\documentclass[aoas,MSNbibl,nameyear,seceqn,dvips]{arximspdf}
\usepackage{graphicx}

\doi{10.1214/12-AOAS574} 
\volume{7}
\issue{1}
\pubyear{2013}
\firstpage{102}
\lastpage{125}

\makeatletter
\newtheorem{theorem}{Theorem}
\newproclaim{assumption}{Assumption}
\newcommand{\ve}[1]{\bolds{#1}}
\makeatother

\begin{document}
\begin{frontmatter}

\title{Varying coefficient model for modeling diffusion tensors along white matter tracts\thanksref{T1}}\vspace*{-4pt} 
\runtitle{Varying coefficient model}
\thankstext{T1}{Supported in part by NIH Grants RR025747, CA142538, MH086633, GM70335,
CA74015, HD03110, EB002779, EB00514901, MH064065, HD053000 and MH070890.
The readers are welcome to request reprints from Dr. Hongtu Zhu
(\printead*{e11}).}

\begin{aug}
\author[A]{\fnms{Ying} \snm{Yuan}\ead[label=e0]{ying.yuan@stjude.org}},
\author[B]{\fnms{Hongtu} \snm{Zhu}\corref{}\ead[label=e1]{htzhu@email.unc.edu}\ead[label=e11]{hzhu@bios.unc.edu}},
\author[B]{\fnms{Martin} \snm{Styner}\ead[label=e2]{martin\_styner@ieee.org}},
\author[B]{\fnms{John~H.}~\snm{Gilmore}\ead[label=e3]{John\_Gilmore@med.unc.edu}}
\and
\author[B]{\fnms{J. S.} \snm{Marron}\ead[label=e4]{marron@email.unc.edu}}

\runauthor{Y. Yuan et al.}
\affiliation{St. Jude Children's Research Hospital,
University of~North Carolina at~Chapel Hill,
University of~North Carolina at~Chapel Hill,
University~of~North Carolina at~Chapel Hill
and
University~of~North~Carolina at~Chapel Hill}
\address[A]{Y. Yuan\\
Department of Biostatistics\\
MS 768, Room 6009\\
St. Jude Children's Research Hospital\\
262 Danny Thomas Place\\
Memphis, Tennessee 38105-3678\\
USA\\
\printead{e0}}
\address[B]{H. Zhu\\
M. Styner\\
J. H. Gilmore\\
J. S. Marron\\
University of North Carolina at Chapel Hill\\
3108-B McGavran Greenberg Hall\\
CB\#7420\\
Chapel Hill, North Carolina 27599\\
USA\\
\printead{e1}\\
\phantom{E-mail:}\ \printead*{e2}\\
\phantom{E-mail:}\ \printead*{e3}\\
\phantom{E-mail:}\ \printead*{e4}} 
\end{aug}

\received{\smonth{9} \syear{2011}}
\revised{\smonth{4} \syear{2012}}

%
\begin{abstract}
Diffusion tensor imaging provides important information on tissue
structure and orientation of fiber tracts in brain white matter in
vivo. It results in diffusion tensors, which are $3\times3$ symmetric
positive definite (SPD) matrices, along fiber bundles.
This paper develops a functional data analysis framework to model
diffusion tensors along
fiber tracts as functional data in a Riemannian manifold with
a set of covariates of interest, such as age
and gender.
We propose a statistical model with varying coefficient functions
to characterize the dynamic association between functional SPD
matrix-valued responses
and covariates. We
calculate weighted least squares estimators of the varying coefficient
functions for the log-Euclidean metric in
the space of SPD matrices.
We also develop a global test statistic to
test specific hypotheses about these coefficient functions and
construct their
simultaneous confidence bands.
Simulated data are further used to
examine the finite sample performance of the estimated varying
coefficient functions.
We apply our model to study potential
gender differences and find a statistically significant aspect of
the
development of diffusion tensors along the right
internal capsule tract in a clinical study of neurodevelopment.\vspace*{-3pt}
\end{abstract}

%
\begin{keyword}
\kwd{Confidence band}
\kwd{diffusion tensor imaging}
\kwd{global test statistic}
\kwd{varying coefficient model}
\kwd{log-Euclidean metric}
\kwd{symmetric positive matrix}.\vspace*{-3pt}
\end{keyword}

\end{frontmatter}

\section{Introduction}\label{Introd}
Diffusion Tensor Imaging (DTI), which measures the effective diffusion of
water molecules, can
provide important information on
the microstructure of fiber tracts and the major neural connections in
white matter [\citeauthor{Basser1994a} (\citeyear{Basser1994b,Basser1994a})]. It has been widely used
to assess the integrity of
anatomical connectivity in white matter.
In DTI, a $3\times3$ symmetric positive definite (SPD) matrix, called
a diffusion tensor (DT), and its three eigenvalue-eigenvector pairs $\{
(\lambda_k, \mathbf{v}_k)\dvtx  k=1, 2, 3\}$ with $\lambda_1\geq\lambda_2\geq
\lambda_3$ are estimated to quantify the degree of diffusivity and the
directional dependence of water diffusion in each voxel (volume pixel).
Multiple fiber tracts in white matter can be constructed by
consecutively connecting the estimated principal directions ($\mathbf{v}_1$) of the estimated DTs in adjacent voxels [\citet{Basser2000}].
Subsequently, some tensor-derived scalar quantities, such as fractional
anisotropy (FA) and mean diffusivity (MD), are commonly estimated along
these white matter fiber tracts for each subject. Specifically,
$\operatorname{MD}=(\lambda_1+\lambda_2+\lambda_3)/3$ describes the amount of
diffusion, whereas FA describes the relative degree of anisotropy and
is given by
%
\begin{equation}
\label{FAeqn} \operatorname{FA}=\sqrt{\frac{3\{(\lambda_1-\bar{\lambda})^2+(\lambda_2-\bar
{\lambda})^2
+(\lambda_3-\bar{\lambda})^2\}}{2(\lambda_1^2+\lambda_2^2+\lambda_3^2)}}.
\end{equation}

In the recent DTI literature, there is an extensive interest in
developing fiber-tract
based analysis for comparing DTIs in population studies [\citet{Goldsmith2011},
\citet{Goodlett2009},
\citet{ODonnellWestin2009},
\citet{Smith2006},
\citet{Yushkevich2008},
\citeauthor{Zhu2010} (\citeyear{Zhu2010,zhu2011})].
The reason is that the region-of-interest (ROI) method primarily
computes averages
of diffusion properties in some manually drawn ROIs, generates various
summary statistics per
ROI, and then carries out statistical analysis on these summary statistics.
This method suffers from identifying meaningful ROIs, particularly the
long curved structures common in
fiber tracts, the instability of statistical results obtained from ROI
analysis, and the partial
volume effect in relative large ROIs [\citet{zhu2011}].
The fiber-tract based analysis usually consists of
two major components, including DTI atlas building and a follow-up
statistical analysis [\citet{Goodlett2009},
\citet{Smith2006},
\citet{Zhu2010}].
The DTI atlas building is primarily to extract DTI
fibers and to establish DTI fiber correspondence across all DTI
data sets from different subjects.
The key steps of the DTI atlas building include DTI registration,
atlas fiber tractography and fiber
parametrization.
Finally, we get a set of individual tracts with the same corresponding
geometry but varying DTs and diffusion properties.
Some statistical approaches have been developed for the analysis of
scalar tensor-derived quantities along fiber tracts
[\citet{Goldsmith2011},
\citet{Goodlett2009},
\citet{Smith2006},
\citet{Yushkevich2008},
\citet{ZhuFMVCMmath2010},
\citeauthor{Zhu2010} (\citeyear{Zhu2010,zhu2011})], but
little has been done on the analysis of whole DTs along fiber tracts,
which is the focus of this paper.


There is a growing interest in the DTI literature in developing
statistical methods for the direct analysis of
DTs in the space of SPD matrices [\citet{Dryden2009}]. \citet
{Schwartzman2008} proposed parametric models for analyzing SPD matrices and
derived the distributions of several test statistics for comparing
differences between the
means of the two (or multiple) groups of SPD matrices.
\citet{KimRichards2010} developed a nonparametric estimator of the
density function of
a random sample of SPD matrices. \citet{Zhu2009} developed a
semiparametric regression model with
SPD matrices as responses and covariates in a Euclidean space.
\citet{Barmpoutis2007} and \citet{Davis2007} developed nonparametric
methods, including
tensor spline methods and local constant regression, to interpolate
diffusion tensor fields.
However, no one has ever developed statistical methods for functional
analysis of DTs along fiber tracts.

In this paper, we propose a \textit{varying coefficient model for
DT-valued functions} (VCDF). We use
varying coefficient functions to characterize the varying association
between diffusion
tensors along fiber tracts and a set of covariates. Here, the varying
coefficients are the parameters in the model which vary with location.
Since the impacts of the covariates of interest may vary spatially, it
would be more sensible to treat the covariates as functions of location
instead of constants, which leads to varying coefficients.
In addition, we explicitly model
the within-subject correlation among multiple DTs measured along a
fiber tract for each subject.
To account for the curved nature of the SPD space, we employ the
log-Euclidean framework
in \citet{Arsigny2006} and then use a weighted least squares
estimation method to estimate the varying coefficient functions.
We also develop a global test statistic to test hypotheses on the
varying coefficient functions and use a resampling method to approximate
the $p$-value.
Finally, we construct a simultaneous confidence band to quantify the
uncertainty of each estimated coefficient function and
propose a resampling method to approximate its critical points.
To the best of our knowledge, this is the first paper for developing
a statistical framework for modeling functional manifold-valued
responses with covariates in Euclidean space.

There are several advantages of the analysis of DTs over the analysis
of scalar diffusion properties along fiber tracts.
The first one is that it can avoid the statistical artifacts, including
biased parameter estimates and incorrect test statistics and $p$-values for
hypotheses of interest, created by comparing the biased
diffusion properties along fiber bundles. This is because the real DT
data estimated from the diffusion weighted images (DWIs) using
weighted least squared methods
are almost unbiased [\citet{Zhu2007b}], whereas the diffusion
properties, which are nonlinear and linear functions
of three eigenvalues of DT data, may be substantially different
from the true diffusion properties [\citet{Anderson2001},
\citet{Pierpaoli1996},
\citet{Zhu2007b}].
In addition, as shown in \citet{Ying2011A}, directly modeling DTs along
fiber bundles as a smooth SPD process allows us to incorporate
a smoothness constraint to further reduce noise in the estimated DTs
along the fiber bundles. This leads to the further reduction of noise in
estimated scalar diffusion properties along the fiber bundles and less
biased estimators of diffusion properties as shown in Figure \ref
{figBias} in Section~\ref{Simulation}.
Moreover, the sole use of diffusion properties, which ignores the
directional information of DT, can decrease the statistical power in
detecting the difference in DTs oriented in different directions.

The rest of the paper is organized as follows.
Section~\ref{Method} presents VCDF and related statistical inference.
Section~\ref{Simulation} examines the finite sample performance of
VCDF via a simulation study.
Section~\ref{RealData} illustrates
an application of VCDF in a clinical study of neurodevelopment.
Section~\ref{discussion} presents concluding remarks.


\section{Data and methods} \label{Method}


\subsection{Early brain development study of white matter tracts}\label{data}

We consider 96 healthy infants ($36$ males and $60$ females)
from the neonatal project on early brain
development led by Dr. Gilmore at the University of North Carolina
at Chapel Hill.
The mean gestational age of these infants
is 245.6 days with SD: 18.5 days (range: 192--270 days).
A 3T Allegra head only MR system was used to acquire all the images.
The system was
equipped with a maximal gradient strength of 40~mT$/$m and a maximal
slew rate of 400 mT$/$(m${}\cdot{}$msec). The DTIs were obtained by
using a single shot EPI DTI sequence (TR/TE${}=5400/73$ msec) with eddy
current compensation. The six noncollinear directions at
the $b$-value of 1000 s$/$mm$^2$ with a
reference scan ($b=0$) were applied. The voxel resolution was
isotropic 2 mm, and the in-plane field of view was set to 256 mm in both
directions. To improve the signal-to-noise ratio of the DTIs, a
total of five repetitions were acquired and averaged.

\begin{figure}

\includegraphics{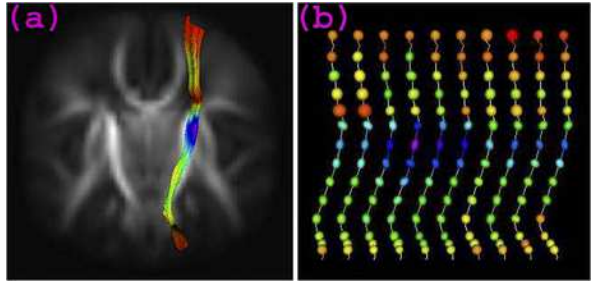}

\caption{\textup{(a)} The fiber bundle of the right internal capsule
fiber tracts in the atlas space.
\textup{(b)} The ellipsoidal representations of full tensors along a
representative right internal capsule fiber tract obtained from each of
$10$ selected
subjects, colored with fractional anisotropy (FA) values. The rainbow
color scheme is used with red corresponding to low FA value and purple
corresponding to high FA value.}
\label{figRIC}
\end{figure}

We processed the DTI data set as follows. We used a weighted least squares
estimation method [\citet{Basser1994b},
\citet{Ying2008},
\citet{Zhu2007b}] to construct
the diffusion tensors.
We used a DTI atlas building pipeline [\citet{Goodlett2009},
\citet{Zhu2010}]
to register DTIs from multiple subjects to create
a study-specific unbiased DTI atlas, to track fiber tracts in the
atlas space,
and to propagate them back into each subject's native space by using
registration information.
Then, we calculated DTs and their scalar diffusion properties at each
location along each individual fiber tract by using
DTs in neighboring voxels close to the fiber tract.
Since the description of the DTI atlas building has been
described in detail [\citet{Goodlett2009},
\citet{Zhu2010}], we do not include
these image processing steps here for the sake of simplicity.
Figure~\ref{figRIC}(a) displays the fiber bundle of the right
internal capsule fiber tract (RICFT), which is an area of white matter
in the brain.
The internal capsule, which lies between the lenticular and caudate
nuclei, consists of a group of myelinated fiber tracts including axons
of pyramidal and extrapyramidal upper motor neurons that connect the
cortex to the cell bodies of lower motor neurons. Although the internal
capsule ends within the cerebrum, the axons that pass through it
continue down through brain stem and spinal cord. It was found that
neonatal microstructural development of the internal capsule tract
correlates with severity of gait and motor deficits [\citet{RIC2007}].
Figure~\ref{figRIC}(b) presents DTs along a representative RICFT
obtained from each of $10$ subjects, in which each DT
is geometrically represented by an ellipsoid.
In this ellipsoidal representation, the lengths of the semiaxes of the
ellipsoid equal the square root of the three eigenvalues of a DT, while
the three eigenvectors define the direction of the three axes.

Our final data set includes DTs and diffusion properties sampled along
the RICFT and a set of covariates of interest from all $n=96$ subjects.
Specifically, let $\operatorname{Sym}^+(3)$ be the set of $3\times3$ SPD
matrices and
$x_j\in[0, L_0]$ be the arc length of the $j$th point on the RICFT
relative to a fixed end point for $j=1, \ldots, n_G$, where $L_0$ is
the longest arc length and $n_G$ is the number of points on the RICFT.
For the $i$th subject,
there is a diffusion tensor at the $j$th point on the RICFT, denoted by
$S_{i}(x_j)\in\operatorname{Sym}^+(3)$, for $i=1, \ldots, n$. Let $\mathbf{z}_i$
be an $r\times1$ vector of covariates of interest.
In this study, we have two specific aims. The first one is to compare
DTs along the RICFT between the male and female groups. The second one
is to delineate the development of fiber DTs across time, which is
addressed by including the gestational age at MRI scanning as a covariate.
Finally, our real data set can be represented as $\{(\mathbf{z}_i; (x_1,
S_i(x_1)), \ldots, (x_{n_G}, S_i(x_{n_G})))\dvtx  i=1, \ldots, n\}$.

\subsection{Varying coefficient model for SPD matrix-valued functional
data}\label{model}
In this section we present
our VCDF.
The code for VCDF written in Matlab along with
its documentation and a sample data\vadjust{\goodbreak} set will be freely accessible from
\url{http://www.bios.unc.edu/research/bias/software.html}.
To make the code easily accessible, we developed a
Graphical User Interface (GUI), also freely
downloadable from the same website.

To proceed, we need to introduce some notation. Let $\operatorname{Sym}(3)$ be
the set of $3\times3$ symmetric matrices with real entries.
For any $A=(a_{kl})\in\operatorname{Sym}(3)$, we define
$\operatorname{vecs}(A)=(a_{11}, a_{2 1}, a_{2 2}, a_{31}, a_{32}, a_{33})^T$
to be a $6 \times1$ vector and
\[
\operatorname{vec}(A)=(a_{11}, a_{12}, a_{13},
a_{21}, a_{22}, a_{23}, a_{31},
a_{32}, a_{33})^T
\]
to be a $9\times1$ vector.
Let $\operatorname{Ivecs}(\cdot)$ be the inverse operator of $\operatorname{vecs}(\cdot
)$ such that $\operatorname{Ivecs}(\operatorname{vecs}(A))=A$ for any $A\in\operatorname{Sym}(3)$.
The matrix exponential of $A\in\operatorname{Sym}(3)$ is given by $\operatorname
{exp}(A)=\sum_{m=0}^{\infty} A^m/m!\in\operatorname{Sym}^+(3)$. For any
$3\times3$ SPD matrix $S$, there is a logarithmic map of $S$, denoted
as $\log(S)=A\in\operatorname{Sym}(3)$, such that $\operatorname{exp}(A)=S$. Let $\mathbf{a}^{\otimes2}=\mathbf{aa}^T$
for any vector or matrix $\mathbf{a}$.

Since the space of SPD matrices is a curved space, we use the
log-Euclidean metric [\citet{Arsigny2006}] to account for the curved
nature of the SPD space. Specifically, we take the logarithmic map of
the DTs $S_i(x)\in\operatorname{Sym}^+(3)$ to get $\log(S_i(x))
\in\operatorname{Sym}(3)$, which has the same effective dimensionality as a
six-dimensional Euclidean space. Thus, we only model the lower
triangular portion of
$\log(S_i(x))$ as follows:
%
\begin{equation}
\label{MVCMeq0}
\operatorname{vecs}\bigl(\log\bigl(S_i(x)\bigr)\bigr)=
{B}(x) \mathbf{z}_i+\mathbf{u}_{i}(x)+{\ve\varepsilon
}_{i}(x),
\end{equation}
where ${B}(x)$ is a $6\times r$ matrix of varying coefficient functions
for characterizing the dynamic associations between
$S_i(x)$ and $\mathbf{z}_i$, $\mathbf{u}_{i}(x)$ is a $6\times1 $ vector
characterizing the within-subject correlation between the
log-transformed DTs,
and ${\ve\varepsilon}_i(x) $ is a $6\times1$ vector of measurement errors.
It is also assumed that ${\ve\varepsilon}_i(x)$ and $\mathbf{u}_i(x)$ are
independent and identical copies of $\operatorname{SP}(\mathbf{{0}}, \Sigma_{{\ve\varepsilon
}})$ and $\operatorname{SP}(\mathbf{{0}}, \Sigma_{\mathbf{u}}),$ respectively,
where $\operatorname{SP}(\mathbf{0}, \Sigma)$ denotes a stochastic process with
mean $\mathbf{0}$ and covariance matrix function $\Sigma(x, x')$ for any
$x, x' \in[0,L_0].$
Let $\mathbf{1}(\cdot)$ be an indicator function.
Assume that ${\ve\varepsilon}_i(x)$ and ${\ve\varepsilon}_i(x')$ for $x\not
=x'$ are independent and, thus, $\Sigma_{{\ve\varepsilon}}(x, x')=\Sigma_{{\ve\varepsilon}} (x, x)\mathbf{{1}}({x=x'})$.
It follows that the covariance structure of $ \operatorname{vecs}(\log
(S_i(x_j))),$ denoted by $\Sigma_S(x, x'),$ is given by
%
\begin{equation}
\label{MVCMeq2} \Sigma_S\bigl(x, x'\bigr)=
\Sigma_\mathbf{u}\bigl(x,x'\bigr)+\Sigma_{\ve\varepsilon}(x,x){
\mathbf{1}}\bigl({x=x'}\bigr).
\end{equation}
Model (\ref{MVCMeq0}) is a multivariate varying coefficient model with
a $6\times1$ vector response and, thus, it
can be regarded as a generalization of univariate varying coefficient
models, which have been widely studied and developed for longitudinal,
time series and functional data [\citet{MR1959093},
\citeauthor{MR1742497} (\citeyear{MR1742497,MR2425354})
\citet{MR2504204},
\citet{MR1769751}].

\subsection{Weighted least squares estimation}\label{betaEst}
Before estimating the varying coefficient functions in $B(x)$, we need
to introduce a few facts about the log-Euclidean metric for the space
of SPDs [\citet{Arsigny2006}].
The use of the log-Euclidean metric results in classical Euclidean
computations in the domain of matrix logarithms.\vadjust{\goodbreak}
Particularly, under the log-Euclidean metric, the geodesic distance
between $S_1$ and $S_2$ in $\operatorname{Sym}^+(3)$ is uniquely given by
%
\begin{equation}
\label{LocalSPD29} d(S_1,S_2)=\sqrt{\operatorname{tr}\bigl[
\bigl\{\log(S_1)-\log(S_2)\bigr\}^{\otimes2}\bigr]},
\end{equation}
which equals the Euclidean distance between $\log(S_1)$ and $\log(S_2)$
in Euclidean space $\operatorname{Sym}(3)$.
However, there is a subtle, but important, difference between regarding
$S(x)$ as a single point in $\operatorname{Sym}^+(3)$ and treating $\log(S(x))$
as a vector in Euclidean space.
By regarding $S(x)$ as a point in $\operatorname{Sym}^+(3)$, we treat all
elements of $S(x)$ as a single unit and use a single bandwidth to
smooth DTs. In contrast,
by treating $\log(S(x))$ as a vector in Euclidean space, traditional
smoothing methods smooth each element of $\log(S(x))$ independently
[\citet{Fan1996},
\citet{Wand1995},
\citet{Wu2006}].

We use
the local linear regression method and the weighted least squares estimation
to estimate $B(x)$ [\citet{Fan1996},
\citet{Ramsay2005},
\citet{Wand1995},
\citet{Welsh2006},
\citet{Wu2006},
\citet{Zhang2007}]. Since the local linear regression method
adapts automatically at the boundary points [\citet{FanGijbels1992}],
it is ideal for dealing with DTs and scalar diffusion properties along
fiber tracts with two ends (see Figure~\ref{figRIC}).
Let $h^{(1)}$ be a given bandwidth, $\dot B(x)=dB(x)/dx$ be a
$6\times r$ matrix, and $I_r$ be the $r\times r$ identity matrix. Using
Taylor's expansion, we
can expand $B(x_j)$ at $x$ to obtain
%
\begin{equation}
\label{MVCMeq4} B(x_j)\approx B(x)+\dot B(x) (x_j-x)
=B_{h^{(1)}}(x)\bigl\{I_r\otimes\mathbf{y}_{h^{(1)}}(x_j-x)
\bigr\},
\end{equation}
where $\mathbf{y}_{h}(x_j-x)=(1, (x_j-x)/h)^T$ and $B_{h^{(1)}}(x)=[B(x),
h^{(1)}\dot B(x)]$ is a $6\times2r$ matrix.
Based on (\ref{MVCMeq0}) and (\ref{MVCMeq4}), $B(x_j)\mathbf{z}_i$ can be
approximated by
$B_{h^{(1)}}(x)\{I_r\otimes\mathbf{y}_{h^{(1)}}(x_j-x)\}\mathbf{z}_i$.
For a fixed
bandwidth $h^{(1)}$, we can calculate
a weighted least squares estimate of $B_{h^{(1)}}(x)$, denoted by $\hat
{B}_{h^{(1)}}(x)=[\hat B(x; h^{(1)}),  h^{(1)}\dot B(x; h^{(1)})]$, by
minimizing an
objective function given by
%
\begin{eqnarray}
\label{MVCMeq5} &&\sum_{i=1}^n \sum
_{j=1}^{n_G} K_{h^{(1)}}({x}_j-{ x})
\nonumber
\\[-8pt]
\\[-8pt]
\nonumber
&&\hspace*{8pt}\qquad{}\times d
\bigl(\log\bigl(S_i(x_j)\bigr), \operatorname{Ivecs}
\bigl(B_{h^{(1)}}(x)\bigl\{I_r\otimes\mathbf{y}_{h^{(1)}}(x_j-x)
\bigr\}\mathbf{z}_i\bigr) \bigr)^2,
\end{eqnarray}
where $K_{h^{(1)}}(\cdot)=K(\cdot/h^{(1)})/h^{(1)}$ is rescaling of the kernel
function $K(\cdot)$, such as the
Gaussian or uniform kernel [\citet{Fan1996},
\citet{Wand1995}]. The explicit form
of $\hat{B}(x; h^{(1)})$ can be found in Appendix~\ref{sec8}.



We pool the data from all $n$ subjects and develop a cross-validation
method to select an estimated bandwidth $h^{(1)},$ denoted by $\hat
{h}_{e}^{(1)}$.
Let $\hat{B}(x;\break h^{(1)})^{(-i)}$ be the weighted least squares
estimator of $B(x)$ for the bandwidth $h^{(1)}$ based on the
observations with the $i$th subject excluded.\vadjust{\goodbreak}
We define a cross-validation score, denoted by $\operatorname{CV}_1(h^{(1)})$,
as follows:
%
\begin{equation}
\label{MVCMeq8} \qquad\operatorname{CV}_1\bigl(h^{(1)}
\bigr)=(nn_G)^{-1}\sum_{i=1}^{n}
\sum_{j=1}^{n_G}d\bigl(\log
\bigl(S_i(x_j)\bigr), \operatorname{Ivecs}\bigl(\hat{B}\bigl(x;
h^{(1)}\bigr)^{(-i)}\mathbf{z}_i\bigr)
\bigr)^2.
\end{equation}
We select $\hat h_e^{(1)}$ by minimizing $\operatorname{CV}_1(h^{(1)})$. In
practice, within a given range of $h^{(1)}$, the value of $\hat
h_e^{(1)}$ can be approximated by computing $\operatorname{CV}_1(h^{(1)})$
through a series of~$h^{(1)}$.
Finally, we can calculate a weighted least squares estimate
of $B(x),$ denoted by $\hat{B}_{e}(x)=\hat B(x; \hat{h}_{e}^{(1)}).$


\subsection{Smoothing individual functions and estimating covariance
matrices}\label{covEst}

To simultaneously construct the individual function $\mathbf{u}_{i}(x)$,
we also employ the local linear regression method.
Let $\dot{\mathbf{u}}_{i}(x)=d{\mathbf{u}}_{i}(x)/dx$. Taylor's expansion of
$\mathbf{u}_{i}(x_j)$ at~$x$ gives
%
\begin{equation}
\label{MVCMeq9} \mathbf{u}_i(x_j)\approx
\mathbf{u}_{i}(x)+\dot{\mathbf{u}}_{i}(x)
(x_j-x)=U_{i}(x) \mathbf{y}_{h^{(2)}}(x_j-x),
\end{equation}
where $U_{i}(x)=[\mathbf{u}_{i}(x), h^{(2)}\dot\mathbf{u}_{i}(x)]$ is a
$6\times2$ matrix.
For each fixed $x$ and each bandwidth $h^{(2)}$, the weighted least
square estimator of $U_{i}(x)$, denoted by $\hat U_{i}(x;
h^{(2)})=[\mathbf{u}_{i}(x; h^{(2)}), h^{(2)}\dot\mathbf{u}_{i}(x; h^{(2)})]$,
can be calculated by minimizing an objective function given by
\begin{eqnarray*}
\label{MVCMeq10} \sum_{j=1}^{n_G}
K_{h^{(2)}}({x}_j- x)d\bigl(\log\bigl(S_i(x_j)
\bigr), \operatorname {Ivecs}\bigl(\hat B_e(x_j)
\mathbf{z}_i+ U_{i}(x)\mathbf{y}_{h^{(2)}}(x_j-x)
\bigr)\bigr)^2.
\end{eqnarray*}
Let $R_i$ be an $n_G\times6$ matrix with the $j$th row $\operatorname
{vecs}(\log(S_i(x_j)))-\hat B_e(x_j)\mathbf{z}_i$ and
$\mathcal{S}$ be an $n_G\times n_G$ smoothing matrix with the $(i,j)$th
element $\tilde{K}^{0}_{h^{(2)}}(x_j-x_i,x_i),$
where $\tilde{K}^{0}_{h^{(2)}}(\cdot,\cdot)$ is the empirical
equivalent kernel [\citet{Fan1996}].
It can be shown that
%
\begin{equation}
\label{MVCMeq12} \bigl(\hat\mathbf{u}_i(x_1), \ldots,
\hat\mathbf{u}_i(x_{n_G}) \bigr)^T=\mathcal{S}
R_i.
\end{equation}


We pool the data from all $n$ subjects and select an estimated
bandwidth of $h^{(2)},$ denoted as $\hat{h}_{e}^{(2)}.$
We define a generalized cross-validation score, denoted by $\operatorname
{GCV}(h^{(2)})$, as follows:
%
\begin{equation}
\label{MVCMeq13g} \operatorname{GCV}\bigl(h^{(2)}\bigr)=n^{-1}
\frac{\sum_{i=1}^n \operatorname{tr}\{(R_i-\mathcal
{S}R_i)^{\otimes2}\} }{\{1-n^{-1}\operatorname{tr}(\mathcal{S})\}^2}.
\end{equation}
We select $\hat h_e^{(2)}$ by minimizing $\operatorname{GCV}(h^{(2)})$.
Like the bandwidth selection in Section~\ref{betaEst}, the value of $\hat
h_e^{(2)}$ can be approximated by computing $\operatorname{GCV}(h^{(2)})$
through a series of $h^{(2)}.$
Finally, by substituting $\hat{h}_{e}^{(2)}$ into
(\ref{MVCMeq12}), we can calculate a weighted least squares estimate
of $\mathbf{u}_i(x),$ denoted by $\hat\mathbf{u}_{ i, e}(x).$

After obtaining $\hat\mathbf{u}_{i,e}(x),$ we can estimate the mean
function $\mathbf{u}(x)$ and
the covariance function $\Sigma_\mathbf{u}(x,x').$ Specifically,\vadjust{\goodbreak} we estimate
$\mathbf{u}(x)$ and $\Sigma_\mathbf{u}(x,x')$ by using their empirical
counterparts based on the estimated $\hat\mathbf{u}_{i,e
}(x)$ as follows:
\[
\hat\mathbf{u}_{e}(x)=n^{-1}\sum
_{i=1}^n\hat\mathbf{u}_{i,e}(x)\quad \mbox
{and}\quad \hat{\Sigma}_\mathbf{u}\bigl(x, x'
\bigr)=(n-6)^{-1}\sum_{i=1}^n
\hat\mathbf{u}_{i,e}(x)\hat\mathbf{u}_{i,e}\bigl(x'
\bigr)^T.
\]


We construct a nonparametric estimator of the
covariance matrix $\Sigma_{\ve\varepsilon}(x,x)$ as follows. Let
$\hat{\ve\varepsilon}_i(x_j)=\operatorname{vecs}(\log(S_i(x_j)))- \hat
B_e(x_j)\mathbf{z}_i-\hat\mathbf{u}_{i,e}(x_j)$
be the estimated residuals for $i=1, \ldots, n$ and $j=1, \ldots, n_G$.
We consider the kernel estimate of $\Sigma_{\ve\varepsilon}(x,x)$ given by
%
\begin{equation}
\label{MVCMeq15} \hat{\Sigma}_{\ve\varepsilon}\bigl(x,x; h^{(3)}
\bigr)=(n-6)^{-1}\sum_{i=1}^n\sum
_{j=1}^{n_G} \frac{ K_{h^{(3)}}(x_j-x) \hat{\ve\varepsilon}_i(x_j)^{\otimes2}} {\sum_{j=1}^{n_G} K_{h^{(3)}}(x_j-x)}.
\end{equation}

We pool the data from all $n$ subjects and select an estimated
bandwidth of $h^{(3)},$ denoted as $\hat{h}_{e}^{(3)}.$ Let $\tilde
{\Sigma}_{\ve\varepsilon}(x_j,x_j)=(n-6)^{-1}\sum_{i=1}^n \hat{\ve
\varepsilon}_i(x_j)\hat{\ve\varepsilon}_i(x_j)^T$ be an estimate of
$\Sigma_{\ve\varepsilon}(x_j, x_j)$ and
$\hat{\Sigma}_{\ve\varepsilon}(x,x; h^{(3)})^{(-i)}$ be the
leave-one-out weighted least squares estimator of $\hat{\Sigma}_{\ve
\varepsilon}(x,x)$.
We define a cross-validation score, denoted by $\operatorname{CV}_2(h^{(3)})$,
as follows:
\[
(nn_G)^{-1}\sum_{i=1}^n
\sum_{j=1}^{n_G}\operatorname{tr}\bigl\{\bigl[\hat{\ve
\varepsilon }_i(x_j)^{\otimes2}- \hat{
\Sigma}_{\ve\varepsilon}\bigl(x_j,x_j; h^{(3)}
\bigr)^{(-i)}\bigr]^{\otimes2} \tilde{\Sigma}_{\ve\varepsilon}(x_j,x_j)^{-1}
\bigr\}.
\]
We select $h^{(3)}$ by minimizing $\operatorname{CV}_2(h^{(3)})$.
In practice, within a given range of $h^{(3)}$, the value of $\hat
h_e^{(3)}$ can be approximated by computing $\operatorname{CV}_2(h^{(3)})$
through a series of $h^{(3)}.$
Finally, by substituting $\hat{h}_{e}^{(3)}$ into
(\ref{MVCMeq15}), we can calculate a weighted least squares estimate
of ${\Sigma}_{\ve\varepsilon}(x,x),$ denoted by $\hat{\Sigma}_{{\ve
\varepsilon},e}(x,x).$

\subsection{Asymptotic properties}\label{AsymP}
We will use the following theorems to make statistical inference on
varying coefficient functions.
The detailed assumptions of these theorems can be found in Appendix \ref
{assumptions} and their proofs are similar to
those in \citet{ZhuFMVCMmath2010}. Thus, we omit them for the sake of space.
We need some notation. Let $\ddot B(x)=d^2B(x)/dx^2$ and $G(\mathbf{0},
\Sigma)$ be a Gaussian process with zero mean and
covariance matrix function $\Sigma(x, x')$ for any $x, x'\in[0, L_0]$.

\begin{theorem}\label{th1}
If the assumptions \textup{(\ref{ass1})--(\ref{ass6})} in the
Appendix~\ref{assumptions} hold,~then
\[
\sqrt{n}\bigl\{\operatorname{vec}\bigl(\hat{B}\bigl(x; h^{(1)}\bigr)-B(x) -0.5
u_2 \ddot B(x)h^{(1) 2}\bigl[1+o_p(1)\bigr]
\bigr)\dvtx x\in[0, L_0]\bigr\}\Rightarrow X_B(x),
\]
where $\Rightarrow$ denote weak convergence of a sequence of stochastic
processes,
$X_B(\cdot)$ follows a Gaussian process $G(\mathbf{0}, \Sigma_\mathbf{u}\otimes\Omega_\mathbf{z}^{-1} )$, and
$\Omega_\mathbf{z}$ is the limit of $n^{-1}\sum_{i=1}^n \mathbf{z}_i^{\otimes
2}$ as $n\rightarrow\infty$.
\end{theorem}

Theorem~\ref{th1} establishes weak convergence of $\hat B(x; h^{(1)})$ as a
stochastic process indexed by $x\in[0, L_0]$ and
forms the foundation for constructing a global test statistic and
simultaneous confidence bands for $\{B(x)\dvtx  x\in[0, L_0]\}$.\vadjust{\goodbreak}

\begin{theorem}\label{th2}If the assumptions \textup{(\ref{ass1})--(\ref{ass7})} in the
Appendix~\ref{assumptions} hold, then
\[
\sup_{(x, x')\in[0, L_0]^2}\bigl|\hat\Sigma_\mathbf{u}\bigl(x, x';
h^{(3)}\bigr)-\Sigma_\mathbf{u}\bigl(x, x'\bigr)\bigr|=
o_p(1).
\]
\end{theorem}

Theorem~\ref{th2} shows the uniform convergence of $\hat\Sigma_\mathbf{u}(x, x';
h^{(3)})$. This is
useful for constructing global and local test statistics for testing
the covariate effects.

\subsection{Hypothesis tests}\label{HT}
In neuroimaging studies, many scientific questions of interest require
the comparison of fiber bundle diffusion tensors along
fiber bundles across two (or more) diagnostic groups and the assessment
of the
development of fiber bundle diffusion tensors along time. Such
questions can often be
formulated as linear hypotheses of $B(x)$ as follows:
%
\begin{equation}
\label{MVCMeq17} H_0\dvtx C\operatorname{vec}\bigl(B(x)\bigr)=
\mathbf{b}_0(x) \quad \mbox{for all } x  \quad \mbox {vs.} \quad  H_1
\dvtx {C}\operatorname{vec}\bigl(B(x)\bigr) \not=\mathbf{b}_0(x),\hspace*{-35pt}
\end{equation}
where ${ C}$ is a $c\times6r$ matrix of full row rank and $\mathbf{b}_0(x)$ is a given $c\times
1$ vector of functions.

We propose both local and global test statistics. The local test
statistic can identify the exact
location of a significant location on a specific tract. At a given
point~$x_j$ on a specific tract, we
test the local null hypothesis
\[
H_{0}(x_j)\dvtx { C}\operatorname{vec}\bigl(B(x_j)
\bigr)=\mathbf{b}_0(x_j)  \quad\mbox{v.s.}\quad
H_{1}(x_j)\dvtx { C}\operatorname{vec}\bigl(B(x_j)
\bigr)\not=\mathbf{b}_0(x_j).
\]
We use a
local test statistic $T_n(x_j)$ defined by
%
\begin{equation}
\label{MVCMeq18} T_n(x_j)=n \mathbf{d}(x_j)^T
\bigl\{{ C}\bigl(\hat\Sigma_\mathbf{u}(x_j,
x_j)\otimes \hat\Omega_\mathbf{z}^{-1}\bigr){
C}^T\bigr\}^{-1} \,\mathbf{d}(x_j),
\end{equation}
where $\hat\Omega_\mathbf{z}=n^{-1}\sum_{i=1}^n \mathbf{z}_i^{\otimes2}$ and
$\mathbf{d}(x)= { C}\operatorname{vec}(\hat{B}_e(x)-\operatorname{bias}(\hat
B_e(x)))-\mathbf{b}_0(x)$.
Following \citet{MR1804172},\vspace*{1pt} a smaller bandwidth leads to a smaller
value of $\operatorname{bias}(\hat B_e(x))$. Moreover, according to our
simulation studies below, we have found that the effect of dropping
$\operatorname{bias}(\hat{B}_e(x))$ is negligible and, therefore, we drop it
from now on.

To test the null hypothesis $H_{0}\dvtx
{C}\operatorname{vecs}(B(x))=\mathbf{b}_0(x)$ for all $x$, we propose a global test
statistic $\mathbf{T}_n$ defined by
%
\begin{equation}
\label{MVCMeq19} \mathbf{T}_n=\int_0^{L_0}
T_n(x) \,dx.
\end{equation}
Let $G_{C}(\cdot)$ be a Gaussian process with zero mean and covariance
matrix function $\Sigma_{C}(x, x')$, which is the limit of
\begin{eqnarray*}
&&\bigl\{{C}\bigl(\hat\Sigma_\mathbf{u}(x, x)\otimes\hat
\Omega_\mathbf{z}^{-1}\bigr){C}^T\bigr
\}^{-1/2} \bigl\{{ C}\bigl(\hat\Sigma_\mathbf{u}\bigl(x,
x'\bigr)\otimes\hat\Omega_\mathbf{z}^{-1}\bigr){
C}^T\bigr\} \\
&&\qquad{}\times \bigl\{{ C}\bigl(\hat\Sigma_\mathbf{u}
\bigl(x', x'\bigr)\otimes\hat\Omega_\mathbf{z}^{-1}
\bigr){ C}^T\bigr\}^{-1/2}.
\end{eqnarray*}
It follows from Theorem~\ref{th1} that
$
\sqrt{n}\{{ C}(\hat\Sigma_\mathbf{u}(x, x)\otimes\hat\Omega_\mathbf{z}^{-1}){ C}^T\}^{-1/2}
\,\mathbf{d}(x) $ converges weakly to $G_{C}(x).$
Therefore, it follows from the continuous mapping theorem that as both
$n$ and $n_G$ converge to infinity, we have
%
\begin{equation}
\label{MVCMeq19a}
\mathbf{T}_n\Rightarrow\int_0^{L_0}G_{ C}(x)^TG_{C}(x)
\,dx.
\end{equation}
Based on the result (\ref{MVCMeq19a}), we develop a wild bootstrap
method to
approximate the $p$-value of $\mathbf{T}_n$. The detailed steps of the
wild bootstrap method are given in Appendix~\ref{sec7}.

\subsection{Confidence band}\label{SimB}
Based on model (\ref{MVCMeq20}), we construct a confidence band for $S(
{B}(x), \mathbf{z})=\exp(\operatorname{Ivecs}(B(x)\mathbf{z}))\in\operatorname{Sym}^+(3)$
over $x\in[0, L_0]$ for a fixed $\mathbf{z}$. Specifically, at a given
significance level
$\alpha$, we construct a simultaneous confidence region in the space of
SPD matrices for each $\mathbf{z}$ based on the critical value $C_z(\alpha
)$ such that
%
\begin{equation}
\quad P\bigl( d\bigl(S\bigl(B(x), \mathbf{z}\bigr), S\bigl(\hat B(x),
\mathbf{z}\bigr)\bigr)\leq C_{z}(\alpha)  \mbox{ for all }  x\in[0,
L_0]\bigr)=1-\alpha.
\end{equation}
Note that
$d(S(B(x), \mathbf{z}), S(\hat B(x), \mathbf{z}))=\sqrt{\operatorname{tr}( [\operatorname
{Ivecs}(\{\hat B_e(x)-B(x)\}\mathbf{z})]^{\otimes2}) }.$
By using Theorem~\ref{th1}, we have that as $n\rightarrow\infty$,
\[
\sqrt{n}d\bigl(S\bigl(B(x), \mathbf{z}\bigr), S\bigl(\hat B(x), \mathbf{z}\bigr)
\bigr)\Rightarrow\sqrt {\operatorname{tr}\bigl[\bigl\{\operatorname{Ivecs}\bigl(X_B(x)
\mathbf{z}\bigr)\bigr\}^{\otimes2}\bigr] }.
\]
We develop an efficient resampling method [\citet{Kosorok2003},
\citet{Zhu2007a}] to approximately draw random samples from $\{X_B(x)\dvtx  x\in
[0, L_0]\}$, denoted by
$\{X_B(x)^{(g)}\dvtx  x\in[0, L_0]\}$ for $g=1, \ldots, G$. The detailed
steps of such a resampling method can be found in Appendix~\ref{sec8}.
Subsequently, we can
calculate $\sqrt{\operatorname{tr}[\{\operatorname{Ivecs}(X_B(x)^{(g)}\mathbf{z})\}^{\otimes2}] }$ for all $g$ and use them to
approximate $C_{z}(\alpha)$ for any given $\alpha$.

Moreover, for $B(x)=(\beta_{kl}(x))$, we can construct confidence bands
for its individual varying coefficient function $\beta_{kl}(x)$ for all
$(k,l)$, $k=1,\ldots, 6$ and $l=1,\ldots, r$.
Specifically, at a given significance level
$\alpha$, we construct a confidence band for each ${\beta}_{kl}(x)$
such that
%
\begin{equation}
\label{MVCMeq20} P\bigl( \hat{\beta}^{L, \alpha}_{kl}(x)<{
\beta}_{kl}(x)<\hat{\beta}^{U,
\alpha}_{kl}(x)  \mbox{ for
all }  x\in[0, L_0]\bigr)=1-\alpha,
\end{equation}
where $\hat{\beta}^{L, \alpha}_{kl}(x)$ and $\hat{\beta}^{U, \alpha
}_{kl}(x)$ are the lower and upper limits of the confidence band.
Let $\mathbf{e}_{kl}$ be a $6r\times1$ vector with the $(l-1)r+k$th
element equal to 1 and all others equal to 0. It follows from Theorem~\ref{th1}
and the continuous mapping theorem that
\[
\sup_{x\in[0, L_0]}\bigl|\sqrt{n}\bigl\{\hat{\beta}_{kl, e}(x)-
\beta_{kl}(x) \bigr\} \bigr|\Rightarrow\sup_{x\in[0, L_0]}\bigl|\mathbf{e}_{kl}^TX_B(x)\bigr|.
\]
We define the critical point $C_{kl}(\alpha)$ to satisfy $P( \sup_{x\in
[0, L_0]}|\mathbf{e}_{kl}^TX_B(x)|\leq C_{kl}(\alpha))=1-\alpha$. Thus,
a $1-\alpha$ simultaneous confidence band for $\beta_{kl}(x)$ is given by
%
\begin{equation}
\label{MVCMeq21} \biggl( \hat{\beta}_{kl, e}(x) -\frac{C_{kl}(\alpha)}{\sqrt{n}},
\hat{\beta}_{kl, e}(x) +\frac{C_{kl}(\alpha)}{\sqrt{n}} \biggr).
\end{equation}
Similar to $C_z(\alpha)$,
the critical point $C_{kl}(\alpha)$ can be approximated as the $1-\alpha
$ empirical percentile of $\sup_{x\in[0, L_0]}|\mathbf{e}_{kl}^TX_B(x)^{(g)}|$ for all $g=1, \ldots, G$.

\section{Simulation study} \label{Simulation}
\begin{figure}

\includegraphics{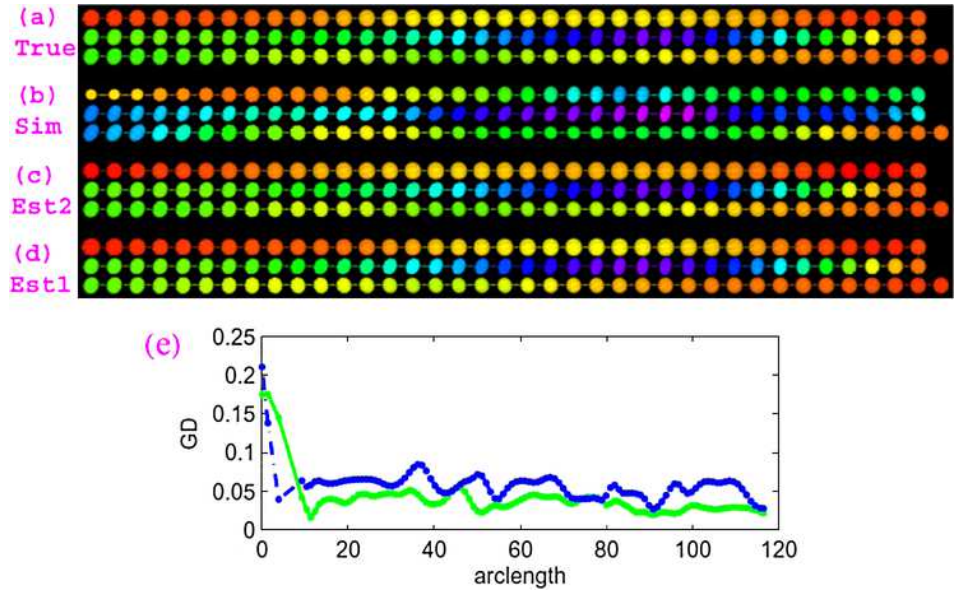}

\caption{ Ellipsoidal representations of the true \textup{(a)}, simulated \textup{(b)}
and estimated \textup{(c)} (based on three covariates) and \textup{(d)} (based on two
covariates) diffusion tensors along the RICFT, colored with their FA
values. The rainbow color scheme is used with red corresponding to low
FA value and purple corresponding to high FA value.
Each set of $3$ rows in \textup{(a)--(d)} represents one tract of $112$ DTs and
the three rows are read from left to right in the top row, right to
left in the middle row and then left to right in the bottom row.
\textup{(e)} Mean geodesic distances between the estimated and true diffusion
tensors (green solid line based on three covariates and blue
dash-dotted line based on two covariates) along the RICFT.}
\label{figEstDT}
\end{figure}

\begin{figure}

\includegraphics{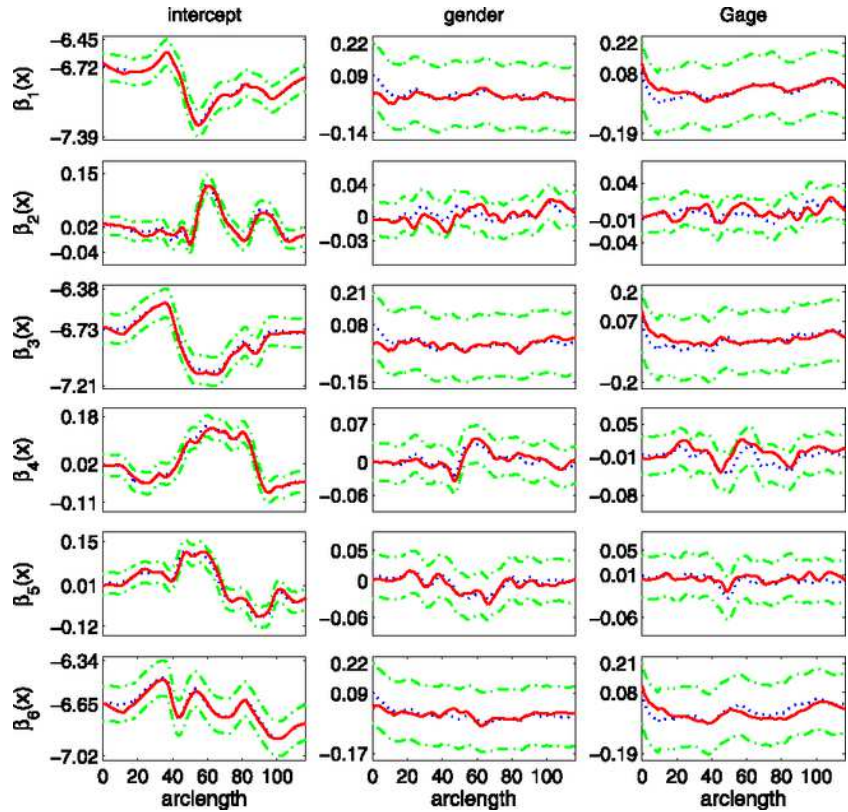}

\caption{Typical $95\%$ simultaneous confidence bands for varying coefficient
functions $\beta_{kl}(x)$.
The solid, dotted and dash-dotted curves are, respectively, the true
curves, the estimated varying coefficient functions and their $95\%$
confidence bands.}
\label{figTypicalSimb5}
\end{figure}

We conducted a Monte Carlo simulation study to examine the finite
sample performance of VCDF.
At each point $x_j$ along the RICFT, the noisy diffusion tensors are
simulated according to the following model:
%
\begin{equation}
S_i(x_j)=\exp\bigl(\operatorname{Ivecs}\bigl(
B(x_j)\mathbf{z}_i+\tau_i\hat
\mathbf{u}_i(x_j)+\tau_i(x_j)
\hat{\ve\varepsilon}_{i}(x_j)\bigr)\bigr),
\end{equation}
where $\tau_i$ and $\tau_i(x_j)$ were independently generated
from a $N(0,1)$ random generator for $i=1, \ldots, n$ and $j=1, \ldots
, n_G$.
Specifically, we set $n=96$, $n_G=112$ and $\mathbf{z}_i=(1, \mathrm{G}_i,
\operatorname{Gage}_i)$ for $i=1, \ldots, 96$, where $\mathrm{G}_i$ and $\operatorname
{Gage}_i$, respectively, denote gender and gestational age.
To mimic real imaging data, we applied our proposed VCDF method to DTs
along the RICFT from all 96 infants in
our clinical data to estimate $B(x)$ by $\hat{B}_e(x)$,
$\mathbf{u}_i(x)$ by $\hat\mathbf{u}_{i,e}(x)$ via
(\ref{MVCMeq12}), and ${\ve\varepsilon}_i(x)$ by $\hat{{\ve\varepsilon
}}_i(x)=\operatorname{vecs}(\log(S_i(x))-\hat B_e(x)\mathbf{z}_i-\hat\mathbf{u}_{i,e}(x))$.
The curves of the varying coefficient functions of $\hat
{B}_e(x)$ are presented in Figure~\ref{figTypicalSimb5}. According to
our real data analysis in Section~\ref{RealData}, the gestational age effect is
significant for our clinical data.
So we fixed all functions in $B(x)$ at
their corresponding functions in $\hat B_e(x)$ except that the third
column of $B(x)$, denoted by
$(\beta_{13}(x), \ldots, \beta_{63}(x))^T$, was set as $c$ times the
third column of $\hat B_e(x)$ where $c$ is set at different values in
order to study the Types I and II error rates of our global test statistic
in testing the gestational age effect. Figure~\ref{figEstDT}(a)
displays the simulated diffusion tensors along the RICFT at $c=1$.


We have five aims in this simulation study. The first aim is to
investigate the consequence of missing an important covariate.
According to our real data analysis in Section~\ref{RealData}, the Gage effect is
significant, whereas the gender effect is not significant.
We fitted two VCDF models, including three-covariate (intercept,
gender and gestational age) and two-covariate (intercept and gender)
models to smooth the DTs along the RICFT, and compare their performance
in reconstructing the true DTs along the RICFT. Note that the
two-covariate model does not include $\operatorname{Gage}_i$ as a covariate.
Figure~\ref{figEstDT} presents the estimated diffusion tensors using
the three-covariate model [Figure~\ref{figEstDT}(c)] and the
two-covariate model [Figure~\ref{figEstDT}(d)]. Inspecting
Figure~\ref{figEstDT}(e) reveals that the three-covariate model leads
a smaller mean geodesic distance between the true and estimated DTs
compared with the two-covariate model. Thus, the three-covariate model
outperforms the two-covariate one in recovering the true DTs along the RICFT.

The second aim is to investigate the finite sample performance of the
global test statistic $\mathbf{T}_n$ based on the whole DT.
In neuroimaging studies, some scientific
questions require the assessment of the development of
diffusion tensors along fiber tracts across time. We formulated the
questions as testing the null hypothesis $H_0\dvtx
\beta_{13}(x)=\cdots=\beta_{63}(x)=0$ for all $x$ along the RICFT. We
first fixed $c=0$ to assess the Type I
error rates for $\mathbf{T}_n$, and then we set
$c=0.2, 0.4, 0.6,$ $0.8$ and $1.0$ to examine the Type II error rates for
$\mathbf{T}_n$ at different effect sizes.

\begin{figure}

\includegraphics{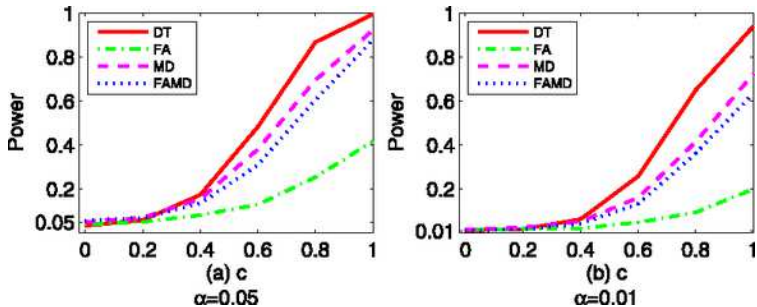}

\caption{Simulation study: Types I and~II error rates as functions
of $c$. Rejection rates
of $T_n$ based on the wild bootstrap method are calculated at six
different values
of the effect size $c$ for sample size $96$ at the \textup{(a)} $0.05$ and \textup{(b)}
$0.01$ significance levels using DTs, FA values, MD values, and joint
values of FA and MD.}
\label{figPower}
\end{figure}

We applied the estimation procedure of VCDF to the noisy DTs along the RICFT.
We approximated
the $p$-value of $\mathbf{T}_n$ by using the wild bootstrap method with $G=1000$
described in Appendix~\ref{sec7}. For each $c$, we set the significance
level~$\alpha$ at both $0.05$ and $0.01$ and used $3000$ replications
to estimate the rejection rate of~$\mathbf{T}_n$. At a fixed $\alpha$, if
the Type I rejection rate is smaller than $\alpha$, then the test is
conservative, whereas if the Type I rejection rate is greater than
$\alpha$, then the test is anticonservative, or liberal.
Figure~\ref{figPower} presents the rejection rates of $\mathbf{T}_n$
across all
effect sizes at the two significance
levels ($\alpha=0.05$ or $0.01$) by using full diffusion tensors. It is
observed that Type I error rates are well maintained
at the two significance levels. In addition, the statistical power for
rejecting the null
hypothesis increases with the effect size and the significance level, which
is consistent with our expectation.

The third aim is to demonstrate the power gain in using DTs compared
with the sole use of diffusion properties.
For each simulated diffusion tensor at $c=0.2, 0.4, 0.6,0.8$ and $1.0$,
we calculated its three eigenvalues $\lambda_1, \lambda_2$ and $\lambda_3$ and
two well-known scalar diffusion properties MD and FA.
To compare the power of our method based on DTs with other methods
based on scalar diffusion properties, we applied
an existing method for the analysis of diffusion properties in \citet
{zhu2011} to three different scenarios: (i) FA, (ii) MD and (iii) (FA, MD).
Then we tested the gestational age effect in each scenario.
Inspecting Figure~\ref{figPower} reveals
that the statistical power for rejecting the null
hypothesis increases with the effect size and the significance level in
all scenarios.
Moreover, compared with the sole use of diffusion properties,
the use of DT dramatically increases the statistical power for
rejecting the null hypothesis.

\begin{figure}

\includegraphics{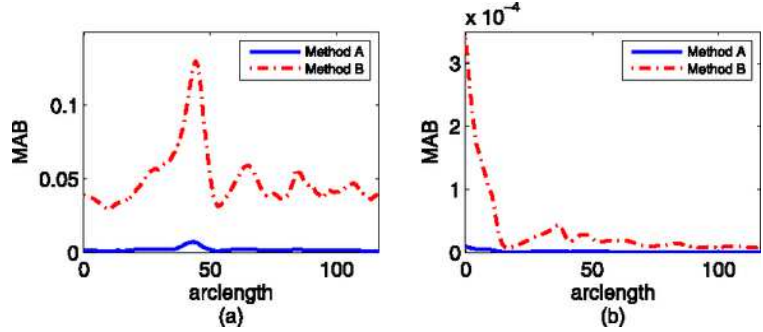}

\caption{Plot of the MAB's of the estimated FA's \textup{(a)} and MD's \textup{(b)} using
two methods A and B
based on 3000 replications. The method A, which uses VCDF by directly
modeling DT's, outperforms
the method B in terms of smaller biases in estimating FA and MD values.}
\label{figBias}
\end{figure}

The fourth aim is to
demonstrate the accuracy gain in estimating scalar diffusion properties
along fiber tracts by directly modeling the DTs using VCDF. We compared
two different methods for estimating
FA's and MD's, here referred to as method A and method B, respectively.
The method A first applies VCDF to
estimate DT's and then calculates the FA or MD curve based on the
estimated DT's. The method B first calculates the FA's or MD's from all
SPD matrices and then uses varying coefficient methods in Euclidean
space to estimate the FA's or MD's. We examined the finite sample
performance of methods A and B by using the Mean Absolute Biases (MAB)
across all $112$ locations, which is defined by
%
\begin{equation}
\label{MVCMeqRB3} \operatorname{MAB}_{Y,j}=96^{-1}\sum
_{i=1}^{96}\Biggl|3000^{-1}\sum
_{s=1}^{3000}\hat {Y}_{sij}-Y_{ij}\Biggr|,
\end{equation}
where $\hat{Y}_{sij}$ is the estimator of $Y_{ij}$, which can be the
estimated FA or MD value at the $j$th location
for the $i$th subject and the $s$th simulation.
Figure~\ref{figBias} reveals that method A has the smaller biases in
estimating FA and MD values and the biases are negligible
compared with those obtained using method B.
This indicates the potential large improvement gained
by directly modeling DT data over method B.

The fifth aim is to
examine the coverage probabilities of the simultaneous confidence bands
for all varying coefficient functions $\beta_{kl}(x)$ in $B(x)$ and
$S(B(x), \mathbf{z})$.
We only considered the generated diffusion tensor data at $c=1$. We
constructed the $95\%$ and $99\%$ simultaneous
confidence bands for all $\beta_{kl}(x)$. Following \citet{MR1804172},
we used a smaller bandwidth with a shrinkage factor $6$ to improve the accuracy
of the confidence bands.

\begin{figure}

\includegraphics{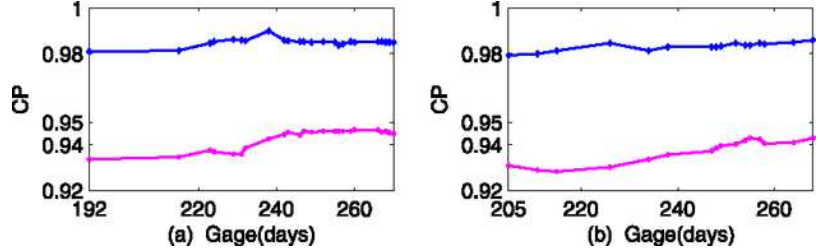}

\caption{Simulated coverage probabilities for $D(\mathbf{z},\beta(x))$
based on 3000 simulations for $\alpha=0.05$ (solid lines with diamond
markers) and $\alpha=0.01$ (solid lines with circle markers),
\textup{(a)} for female \textup{(b)}~for male at different gestational ages, respectively.}
\label{figSimBSPD}
\end{figure}

Table~\ref{Table1} summarizes the empirical coverage probabilities based on 3000
replications for $\alpha=0.01$
and $0.05$. The coverage probabilities are
quite close to the prespecified confidence levels.
Figure~\ref{figTypicalSimb5} presents typical critical values of
$95\%$ simultaneous confidence regions for vectors of coefficient
functions $\beta_{k\cdot}(x)=(\beta_{k1}, \ldots, \beta_{kr})^T,  k=1,
\ldots, 6$.
Figure~\ref{figSimBSPD} summarizes the empirical coverage
probabilities for $S(B(x), \mathbf{z})$ based on 3000 replications at
$\alpha=0.01$
and $0.05$. The coverage probabilities are
quite close to the expected confidence levels.

\begin{table}
\caption{Simulated coverage probabilities for varying coefficient
functions in $B(x)=(\beta_{kl}(x))$ based on 3000 replications at the
significance levels
$\alpha=0.01$ and $0.05$}\label{Table1}
\begin{tabular*}{\textwidth}{@{\extracolsep{\fill}}lcccccc@{}}
\hline
& \multicolumn{3}{c}{$\bolds{\alpha=0.05}$} & \multicolumn{3}{c@{}}{$\bolds{\alpha=0.01}$}
\\[-4pt]
& \multicolumn{3}{c}{\hrulefill} & \multicolumn{3}{c@{}}{\hrulefill} \\
& \textbf{Intercept} & \textbf{Gender} & \textbf{Gage} & \textbf{Intercept} & \textbf{Gender} & \multicolumn{1}{c@{}}{\textbf{Gage}} \\
& $\bolds{l=1}$ & $\bolds{l=2}$ & $\bolds{l=3}$ &$\bolds{l=1}$ & $\bolds{l=2}$ & $\bolds{l=3}$ \\
\hline
$\beta_{1l}(x)$ &0.9497 & 0.9420 & 0.9387 & 0.9867 & 0.9837 & 0.9810 \\
$\beta_{2l}(x)$ &0.9440 & 0.9443 & 0.9383 & 0.9843 & 0.9907 & 0.9857 \\
$\beta_{3l}(x)$ &0.9457 & 0.9383 & 0.9400 & 0.9870 & 0.9833 & 0.9807 \\
$\beta_{4l}(x)$ &0.9480 & 0.9457 & 0.9400 & 0.9880 & 0.9870 & 0.9850 \\
$\beta_{5l}(x)$ &0.9437 & 0.9350 & 0.9350 & 0.9870 & 0.9873 & 0.9823 \\
$\beta_{6l}(x)$ &0.9473 & 0.9400 & 0.9403 & 0.9860 & 0.9827 & 0.9797 \\
\hline
\end{tabular*}
\end{table}

\section{Analysis of the right internal capsule fiber tract} \label{RealData}
We have two specific aims for the analysis of the right internal
capsule fiber tracts. The first one is to compare DTs along the RICFT\vadjust{\goodbreak}
between the male and female groups. The second one is to delineate the
development of fiber DTs across time. To achieve these two aims,
we fitted VCDF to DTs along the RICFT with gestational age at MRI
scanning and gender as covariates.
We applied the estimation procedure in Section~\ref{Method} to estimate
$B(x)$, $\Sigma_\mathbf{u}(\cdot, \cdot)$ and
$\Sigma_{\ve\varepsilon}(\cdot, \cdot)$.
Then, we constructed the
global test statistics $\mathbf{T}_n$ and the local test statistics
$T_n(x_j)$ to test
the gender effect and the gestational age effect based on DTs along the
RICFT. The $p$ value of
$\mathbf{T}_n$ was approximated by using the resampling method with
$G=5000$ replications.
Finally, we constructed the $95\%$
simultaneous confidence bands for the varying coefficient functions
$\beta_{kl}(x)$.\looseness=-1

\begin{figure}

\includegraphics{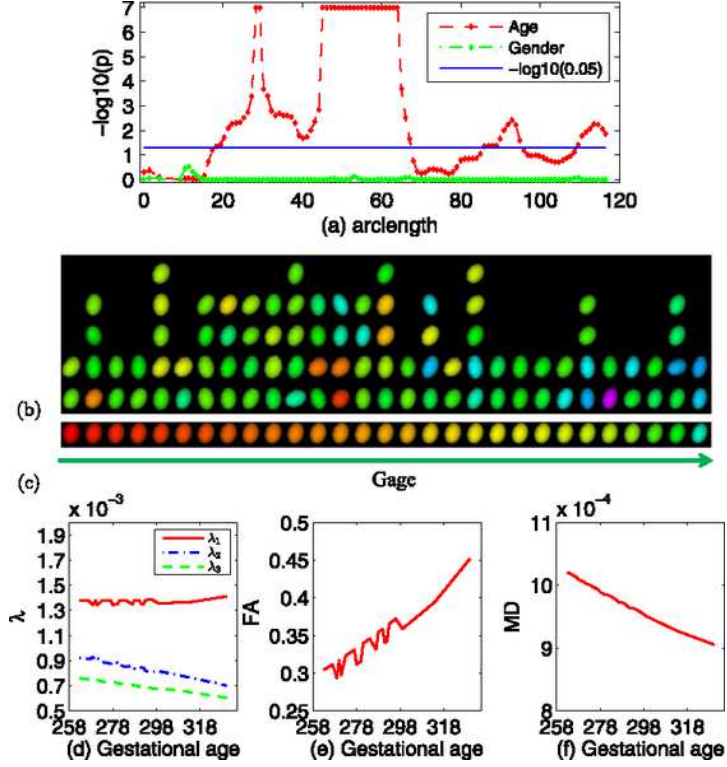}

\caption{\textup{(a)} The $-\log_{10}(p)$ values of test statistics $T_n(x_j)$
for testing gender or gestational age effect of diffusion tensors on
the right internal capsule tract, which shows no significant gender
effect and significant gestational age effect.The ellipsoidal
representations of \textup{(b)} raw and \textup{(c)} smoothed diffusion tensors changing
with the gestational age at one location (at $\operatorname{arclength}=61.02$) on the
right internal capsule tract with significant gestational age effect,
colored with FA values. The rainbow color scheme is used with red
corresponding to low FA value and purple corresponding to high FA
value. The plots of three eigenvalues \textup{(d)}, FA \textup{(e)} and MD \textup{(f)} values at
that location.}
\label{figPvalues}
\end{figure}

To test the gender and gestational age effects, we calculated the local
test statistics $T_n(x_j)$ and their
corresponding $p$ values across all points on the RICFT. It is shown in
Figure~\ref{figPvalues}(a)
that most points do not have $-\log_{10}(p)$ values greater than $1.3$
for testing the gender effect.
Then, we also computed the global test statistic $\mathbf{T}_n=797.65$ and
its associated $p$-value $p=0.3934$, indicating no
gender effect. Inspecting Figure~\ref{figPvalues}(b) reveals that
the $-\log_{10}(p)$ values of $T_n(x_j)$ for testing the gestational
age effect are extremely significant in the middle part of the RICFT. The
global gestational age effect was also found to be
highly significant
with $\mathbf{T}_n=5271.7$ and its $p$-value $p<10^{-6}$.
It indicates that DTs along the RICFT
are significantly associated with the gestational age, even though
there is no gender difference among DTs along the RICFT. In order to
investigate the development of DTs across the gestational age, we chose
a location at $\operatorname{arclength}=61.02$ and observed
that the diffusion tensors become anisotropic and their sizes become
smaller as gestational
age increases [Figures~\ref{figPvalues}(b) and (c)]. Recall that the
three eigenvalues of a DT reflect the magnitude of the diffusion of
water molecules along three directions parallel to its three
eigenvectors and that MD reflects the total magnitude of the diffusion
of water molecules. To show the decreasing trend of DT, we also plotted
the curves of all three eigenvalues and MD values in Figures \ref
{figPvalues}(e) and~(g), respectively, both of which explicitly show
that the first eigenvalue does not change much, whereas the second,
third eigenvalues and MD values decrease with the gestational age. In
addition, it is observed from~\ref{figPvalues}(f) that FA increases
with gestational age, which indicates that DTs become more anisotropic
as gestational age increases.

\begin{figure}

\includegraphics{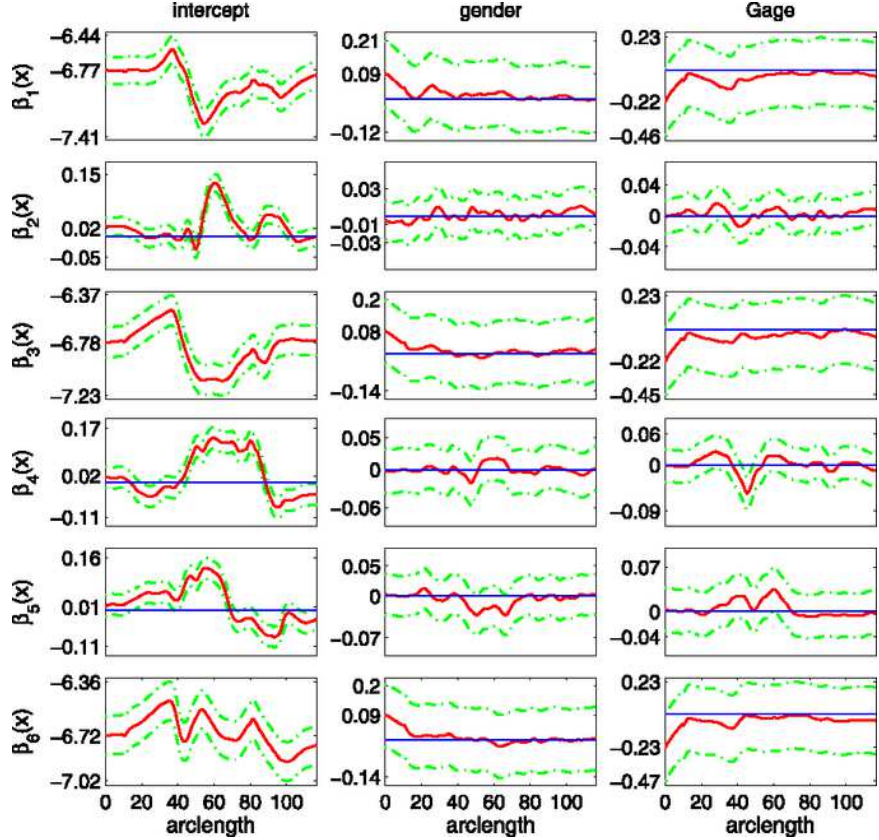}

\caption{$95\%$ simultaneous confidence bands for coefficient
functions. The solid curves are
the estimated coefficient functions and the dashed curves are the $95\%$ confidence bands.
The thin horizontal line is the line crossing the origin.}
\label{figSimb}
\end{figure}

Figure~\ref{figSimb} presents the estimated varying coefficient
functions along with their $95\%$ simultaneous confidence bands. In
Figure~\ref{figSimb} all simultaneous confidence bands contain the
horizontal line crossing $(0,0)$ for the gender effect, whereas the
horizontal line is out of the $95\%$ simultaneous confidence band
for $\beta_{43}(x)$, which indicates the significant gestational age
effect. This agrees with our previous analysis results based on the
global and local test statistics for the gender and gestational age effects.

\begin{figure}

\includegraphics{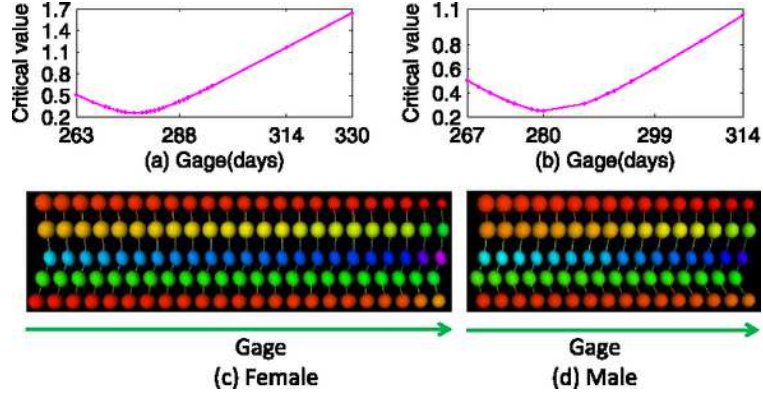}

\caption{The $95\%$ critical values for $S(B(x), \mathbf{z})$ across
gestational ages for female \textup{(a)} and male \textup{(b)} groups, respectively.
The ellipsoidal representation of the estimated $S(B(x), \mathbf{z})$
along the right internal capsule tract across
gestational ages for female \textup{(c)} and male \textup{(d)} groups, respectively,
colored with FA values. The rainbow color scheme is used with red
corresponding to low FA value and purple corresponding to high FA
value. The displayed four rows from the top to the bottom correspond to
DTs at arclength $ 0, 31.22, 61.02, 80.92$ and
$116.47$. Specifically, the third row shows the apparent trend of DT's
changing with gestational age for both
female and male groups.}
\label{figSimBSPDReal}
\end{figure}

Finally, Figure~\ref{figSimBSPDReal} presents the $95\%$ critical
values for $S(B(x), \mathbf{z})$ and the estimated $S(B(x), \mathbf{z})$
along the RICFT across gestational age for female and
male groups, respectively. Inspecting Figure~\ref{figSimBSPDReal}
reveals that the variation of $S(B(x), \mathbf{z})$ is larger on the two
boundary points (especially on the right side)
and smaller in the middle. In addition, the apparent trend of DT's
changing with gestational age is shown at $\operatorname{arc\mbox{-}length}=61.02$ for both
female and male groups.

\section{Discussion}\label{discussion}
In this paper we have developed a
functional data analysis framework, VCDF, for modeling diffusion
tensors along
fibber bundles in the Riemannian manifold of SPD matrices under
the log-Euclidean metric with a set of
covariates of interest. The most important characteristic of our method
is that it is formulated based on the whole diffusion tensors instead
of the DT derived scalar quantities and, thus, it can directly handle
diffusion tensors. In addition, VCDF can characterize the dynamic
association between functional DT-valued responses and covariates by
using a set of varying coefficient functions. Compared with the methods
based on DT derived quantities, such as FA and MD, our method shows the
apparent superiority in estimating DT derived quantities compared with
those based on DT derived quantities (Figure~\ref{figBias}). One
reason is that the
DT data which is estimated from DWIs is almost biased, whereas the DT
derived quantities are linear and nonlinear functions of eigenvalues of
DT data, which are very different from the ground truth. The other
reason is that
directly modeling DTs along fiber bundles as a smooth SPD process
allows us to incorporate
a smoothness constraint to further reduce noise in the estimated
DTs along the fiber bundles. This leads to the further reduction of
noise in
estimated scalar diffusion properties along the fiber bundles.
In addition, our method has the greater statistical power in detecting
the effect of covariates of interest as is shown in Figure \ref
{figPower}. One reason is that VCDF is less biased in parameter
estimation. The other one is that our method accounts for all
information contained in the DTs along the fiber bundles.


Several major issues remain to be addressed in future research.
All fiber-tract-based methods including VCDF are only applicable
to these prominent white matter tracts and do not account for the
uncertainties of tracking these fiber tracts.
It is important to develop new statistical methods to appropriately
account for such uncertainties in fiber-tract analysis especially for
inconspicuous fiber tracts.
VCDF is based on the second-order diffusion tensor. It may be
interesting to extend VCDF to the analysis of high angular
resolution diffusion imaging (HARDI), which is important for resolving
the issue of fiber crossing [\citet{Assemlal2011}].
Furthermore, it would be of great interest to extend VCDF to
longitudinal studies and family studies. Finally,
we have treated DTs along fiber tracts as functional
responses; it would be interesting to treat DTs along fiber tracts as
varying covariate functions to predict a scalar outcome (e.g.,
diagnostic group)
[\citet{Goldsmith2011}].

\begin{appendix}

\section{Assumptions} \label{assumptions}

\renewcommand{\theassumption}{C\arabic{assumption}}
\begin{assumption}\label{ass1} ${\ve\varepsilon}_{i}(x)$ and $\mathbf{u}_{i}(x)$
are identical and independent copies of $\operatorname{SP}(0, \Sigma_{\ve
\varepsilon})$ and $\operatorname{SP}(0, \Sigma_\mathbf{u})$, respectively. ${\ve
\varepsilon}_{i}(x)$ and ${\ve\varepsilon}_{i}(x')$
are independent for any $x\not= x'\in[0, L_0]$.
${\ve\varepsilon}_{i}(x)$ and $\mathbf{u}_{i}(x')$
are independent for any $x, x'\in[0, L_0]$. Moreover, with
probability one, the sample path of $\mathbf{u}_{i}(x)$ has continuous
second-order derivative on $[0, L_0]$ and $E[\sup_{x\in[0, L_0]}\Vert \mathbf{u}_i(x)\Vert_2^{r_1}]<\infty$
and $E\{\sup_{x\in[0, L_0]}[\Vert \dot\mathbf{u}_i(x)\Vert_2+\Vert \ddot\mathbf{u}_i(x)\Vert_2]^{r_2}\}<\infty$ for all $r_1,
r_2\in(2, \infty)$, where $\Vert \cdot\Vert_2$ is the Euclidean norm.
\end{assumption}

\begin{assumption}\label{ass2} All components of $B(x)$ and $\Sigma_{\ve
\varepsilon}(x, x)$ have continuous second-order derivatives on $[0,
L_0]$. The fourth moments of ${\ve\varepsilon}_{i}(x)$ are continuous on
$[0, L_0]$.
All components of $\Sigma_\mathbf{u}(x, x')$ have continuous second-order
partial derivatives with respect to $(x, x')\in[0, L_0]^2$. Moreover,
$\Sigma_{\ve\varepsilon}(x, x)$ and $\Sigma_\mathbf{u}(x,x)$ are positive
for all $x\in[0,
L_0]$.
\end{assumption}

\begin{assumption}\label{ass3} The points ${\mathcal X}=\{x_j, j=1,
\ldots, n_G\}$ are independently and identically distributed with
density function $\pi(x)$, which has the bounded support $[0, L_0]$.
For some constants $\pi_L$ and $\pi_U\in(0, \infty)$ and any $x\in
[0, L_0]$, $\pi_L\leq\pi(x) \leq\pi_U$ and $\pi(x)$ has continuous
second-order derivative.
\end{assumption}

\begin{assumption}\label{ass4} The kernel function $K(t)$ is a
symmetric density function with a compact support [$-1, 1$] and is
Lipschitz continuous.
\end{assumption}

\begin{assumption}\label{ass5} The covariate vectors $\mathbf{z}_i$ are
independently and identically distributed with $E\mathbf{z}_i=\mu_z$ and
$E[\Vert \mathbf{z}_{i}\Vert_2^4]<\infty$ and that $E[\mathbf{z}_{i}^{\otimes2}]
=\Omega_Z$ is invertible.
\end{assumption}

\begin{assumption}\label{ass6} Both $n$ and $n_G$ converge to $\infty$,
$h^{(1)}=o(1)$, $n_Gh^{(1)}\rightarrow\infty$, and
$ h^{(1) -1}|\log h^{(1)}|^{1-2/q_1}\leq n_G^{1-2/q_1}$, where $q_1\in
(2, 4)$.
\end{assumption}

\begin{assumption}\label{ass7}
$E[\Vert {\ve\varepsilon}_{i}(x_j)\Vert_2^{q_2}]<\infty$ for some $q_2\in(4,
\infty)$,
$\max(h^{(2)}, \break h^{(3)})=o(1)$, $n_G(h^{(2)}+h^{(3)})\rightarrow\infty$,
$(h^{(2)})^{-4}(\log n/n)^{1-2/q_2}=o(1)$,
and\break
$ (h^{(3)})^{-2}(\log n/\break n)^{1-2/q_2}=o(1)$.
\end{assumption}

\section{Wild bootstrap method}\label{sec7}

We develop the four key steps of the wild bootstrap method for
approximating the $p$-value of $\mathbf{T}_n$ as follows.
\begin{longlist}
\item[Step (i):] Use the weighted least squares estimation to fit model (\ref
{MVCMeq0}) under the linear constraint specified
in $H_0$, which yields $\hat B_e^*(x_j)$. Calculate $\hat\mathbf{u}_{i,e}^*(x_j)$ according to (\ref{MVCMeq12}) and $\hat{\ve\varepsilon
}_{i,e}^*(x_j)=\operatorname{vecs}(\log(S_i(x_j)))- \hat B_e(x_j)^*\mathbf{z}_i-\hat\mathbf{u}_{i,e}^*(x_j)$ for $i=1, \ldots, n$ and $j=1, \ldots, n_G$.

\item[Step (ii):] Generate a random sample $\tau_i^{(g)}$ and $\tau_i(x_j)^{(g)}$ from a $N(0,
1)$ random generator for $i=1, \ldots, n$ and $j=1, \ldots, n_G$ and
then construct
\[
\hat{S}_i(x_j)^{(g)}=\exp\bigl(\operatorname{Ivecs}
\bigl(\hat B_e^*(x_j)\mathbf{z}_i+
\tau^{(g)}_i\hat\mathbf{u}_{i,e}^*(x_j)+
\tau_i(x_j)^{(g)}\hat{\ve\varepsilon
}_{i,e}^*(x_j)\bigr)\bigr).
\]
Then, based on $\hat{
S}_i(x_j)^{(g)}$, we recalculate $\hat B_e(x)^{(g)}$, and $\mathbf{d}(x)^{(g)}={
C}\hat B_e(x)^{(g)}-\mathbf{b}_0(x)$.
We compute
\begin{eqnarray*}
\mathbf{T}_n^{(g)}&=&\int_0^{L_0}T_n(x)^{(g)}
\,dx,
\\
T_n(x_j)^{(g)}&=&n \,\mathbf{d}(x_j)^{(g)T}
\bigl\{{C}\bigl(\hat\Sigma_\mathbf{u}(x_j, x_j)
\otimes\hat\Omega_\mathbf{z}^{-1}\bigr){C}^T\bigr
\}^{-1}\, \mathbf{d}(x_j)^{(g)}
\nonumber
\end{eqnarray*}
for $j=1, \ldots, n_G.$

\item[Step (iii):] Aggregate the results of Step (ii) over $g=1, \ldots, G$ to
obtain $\{ T_{n,
\max}^{(g)}=\max_{1\leq j\leq n_G}T_n(x_j)^{(g)}\dvtx  g=1,\ldots, G\}$
and calculate $ p(x_j)= G^{-1} \times \sum_{g=1}^G 1(T_{n,
\max}^{(g)}\geq T_n(x_j)) $ for each $x_j$. The $p(x_j)$ is the
corrected $p$-value at the
location $x_j$.

\item[Step (iv):] Aggregate the results of Step (ii) over $g=1, \ldots, G$ to
obtain $\{ \mathbf{T}_n^{(g)}\dvtx
g=1,\ldots, G\}$ and calculate $ p= G^{-1} \sum_{g=1}^G
1(\mathbf{T}_n^{(g)}\geq\mathbf{T}_n)$.\vspace*{1pt} If $p$ is smaller than a prespecified
significance level $\alpha$, say, 0.05, then we reject the null hypothesis
$H_0$.
\end{longlist}
\section{Resampling method for approximating Gaussian process}\label{sec8}

Recall that $B_{h^{(1)}}(x)=[B(x), h^{(1)} \dot B(x)]$ in (\ref
{MVCMeq4}) is a $6\times2r$ matrix. It can be shown that $\hat
B_{h^{(1)}}(x)^T$
is given by
%
\begin{equation}
\label{AppEQ1} \Sigma\bigl(h^{(1)}, x\bigr)^{-1} \sum
_{i=1}^n\sum_{j=1}^{n_G}
K_{h^{(1)}}(x_j-x) \bigl[\mathbf{z}_{i}\otimes
\mathbf{y}_{h^{(1)}}(x_j-x)\bigr] \operatorname{vecs}\bigl(\log
\bigl(S_i(x_j)\bigr) \bigr)^T,\hspace*{-36pt}
\end{equation}
where $\Sigma(h^{(1)}, x)=\sum_{i=1}^n\sum_{j=1}^{n_G}
K_{h^{(1)}}(x_j-x)[\mathbf{z}_{i}^{\otimes2}\otimes\mathbf{y}_{h^{(1)}}(x_j-x)^{\otimes2}]$. Thus,
we can obtain $\hat B(x; h^{(1)})$ as follows:
%
\begin{equation}
\label{AppEQ2} \hat B\bigl(x; h^{(1)}\bigr)=\bigl[{ I}_r
\otimes(1, 0)\bigr] \hat B_{h^{(1)}}(x).
\end{equation}

To approximately simulate from the Gaussian process $X_B(\cdot)$, we
develop a resampling method as follows:
\begin{itemize}
\item Based on $\hat B(x_j; h^{(1)})$, we calculate
$\hat\mathbf{r}_{i}(x_j)=\operatorname{vecs}(\log( S_{i}(x_j)))-\hat B(x_j; h^{(1)})\mathbf{z}_i$ for $i=1, \ldots, n$ and $j=1, \ldots, n_G$.
\item For $g=1, \ldots, G$, we independently simulate $\{\tau_i^{(g)}\dvtx  i=1, \ldots, n\}$ from $N(0, 1)$.
\item For $g=1, \ldots, G$, we calculate a stochastic process
$X_B(x)^{(g)}$ given by
\[
\sqrt{n} \bigl[{ I}_r\otimes(1, 0)\bigr]\Sigma\bigl(h^{(1)},
x\bigr)^{-1} \sum_{i=1}^n
\tau_i^{(g)}\sum_{j=1}^{n_G}
K_{h^{(1)}}(x_j-x) {C}_i\bigl(x_j-x;
h^{(1)}\bigr) \hat\mathbf{r}_{i, l}(x_j)^T,
\]
where ${C}_i(x_j-x; h^{(1)})=[\mathbf{z}_{i}\otimes\mathbf{y}_{h^{(1)}}
(x_j-x)] $ is a $2r\times1$ vector.
\end{itemize}
\end{appendix}


%

%

%


\printaddresses

\end{document}